\documentstyle[12pt]{article}

\newcommand{\bb}{\begin{equation}}
\newcommand{\ee}{\end{equation}}
\newcommand{\bqn}{\begin{eqnarray}}
\newcommand{\eqn}{\end{eqnarray}}
\newcommand{\pp}{\partial}

\begin{document}
\begin{titlepage}

\begin{flushright}

hep-th/9704023\\

\end{flushright}
\vfill

\begin{center}
{\Large{\bf Anomalies and Renormalization of BFYM Theory}}
\end{center}
\vfill

\begin{center}
{\large
Marc Henneaux$^{a,b}$}
\end{center}
\vfill

\begin{center}{\sl
$^a$ Facult\'e des Sciences, Universit\'e Libre de
Bruxelles,\\
Campus Plaine C.P. 231, B--1050 Bruxelles, Belgium\\[1.5ex]

$^b$ Centro de Estudios Cient\'\i ficos de Santiago,\\
Casilla 16443, Santiago 9, Chile

}\end{center}
\vfill

\begin{abstract}
The (perturbative) renormalization properties of the
BF formulation of Yang-Mills gauge models are shown
to be identical to those of the usual, second order formulation.  This result
holds in any number of spacetime dimensions and is a direct consequence
of cohomological theorems established by G. Barnich, F. Brandt and
the author ({\em Commun.Math.Phys.} {\bf 174} (1995) 57).

\end{abstract}

\vspace{5em}

\end{titlepage}

Recently, an interesting  re-interpretation of Yang-Mills theory
\bqn
S_{YM}[A] &=& - \frac{1}{4 g^2} \int d^nx F^a_{\mu \nu}
F_a^{\mu \nu}, \label{YMAction} \\
 F^a_{\mu \nu} &=& \pp_\mu A^a_\nu - \pp_\nu A^a_\mu
 + f^a _{bc} A^b_\mu A^c_\nu \label{FieldStrength}
\eqn
as a deformation of a BF topological theory has been developed
\cite{Fucito,Martellini1,Cattaneo,Martellini2}.
This new interpretation, based on the first order reformulation
of the Yang-Mills action 
\bqn
S_{BFYM}[A, B] &=& 
\int d^n x \epsilon^{\mu_1 \dots \mu_n} B^a_{\mu_1 \dots \mu_{n-2}}
F_{a \mu_{n-1} \mu_n} \nonumber \\
&-&  2 (n-2)! g^2 \int d^n x
B^a_{\mu_1 \dots \mu_{n-2}} B_a^{\mu_1 \dots \mu_{n-2}},
\label{BFYMAction}
\eqn
where $B^a_{\mu_1 \dots \mu_{n-2}}$ is an auxiliary field equal on-shell
to $g^{-2}$ times the dual of the field strength, relies on the
observation that for $g=0$, the action (\ref{BFYMAction}) reduces
to the action of the BF topological system 
\cite{Horowitz,Blau,Horo2,Birmingham},
\bb
S_{top}[A, B] =
\int d^n x \epsilon^{\mu_1 \dots \mu_n} B^a_{\mu_1 \dots \mu_{n-2}}
F_{a \mu_{n-1} \mu_n}. 
\label{TopAction}
\ee

One may exhibit a ``topological symmetry" for (\ref{BFYMAction})
somewhat analogous to the topological symmetry of
(\ref{TopAction}) by introducing a pure gauge field 
$\eta^a_{\mu_1 \dots \mu_{n-3}}$ 
and replacing (\ref{BFYMAction}) by the equivalent
action \cite{Accardi}
\bqn
S_{BFYM\eta}[A, B, \eta ]
 &=&
\int d^n x \epsilon^{\mu_1 \dots \mu_n} B^a_{\mu_1 \dots \mu_{n-2}}
F_{a \mu_{n-1} \mu_n} \nonumber \\
&-&  2 (n-2)! g^2 \int d^n x
{\tilde B}^a_{\mu_1 \dots \mu_{n-2}} {\tilde B}_a^{\mu_1 \dots \mu_{n-2}},
\label{BFYMetaAction}
\eqn
with
\bb
{\tilde B}^a_{\mu_1 \dots \mu_{n-2}} =
B^a_{\mu_1 \dots \mu_{n-2}} + D_{[\mu_1} \eta^a_{\mu_2 \dots
\mu_{n-2}]}.
\label{B}
\ee
While the actions (\ref{YMAction}) and (\ref{BFYMAction}) are
invariant only under the standard Yang-Mills gauge transformations, which
read, in terms of all the variables,
\bb
\delta_\epsilon A^a_\mu = D_\mu \epsilon^a, \;
\delta_\epsilon B^a_{\mu_1 \dots \mu_{n-2}} = f^a_{bc} B^b
_{\mu_1 \dots \mu_{n-2}} \epsilon^c, \;
\delta_\epsilon \eta^a_{\mu_1 \dots \mu_{n-3}} = f^a_{bc}
\eta^b_{\mu_1 \dots \mu_{n-3}} \epsilon^c,
\label{Gauge1}
\ee
the action (\ref{BFYMetaAction}) has the additional ``topological
invariance"
\bb
\delta_\Lambda  A^a_\mu = 0, \;
\delta_\Lambda B^a_{\mu_1 \dots \mu_{n-2}} = -
D_{[\mu_1} \Lambda^a_{\mu_2 \dots \mu_{n-2}]}, \;
\delta_\Lambda \eta^a_{\mu_1 \dots \mu_{n-3}} = 
\Lambda^a_{\mu_1 \dots \mu_{n-3}}.
\label{Gauge2}
\ee
One goes from (\ref{BFYMetaAction}) to  (\ref{BFYMAction})
by gauge-fixing the topological symmetry, e.g., through the
condition $\eta^a_{\mu_1 \dots \mu_{n-3}} = 0$.  One then
recovers the original Yang-Mills action by eliminating the
auxiliary field $B^a_{\mu_1 \dots \mu_{n-2}}$ through
its own equation of motion $\delta S_{BFYM}/\delta
B^a_{\mu_1 \dots \mu_{n-2}} = 0$.  The actions (\ref{YMAction}),
(\ref{BFYMAction}) and (\ref{BFYMetaAction}) are therefore
equivalent.

The BF formulation of Yang-Mills 
theory opens the door to new, quite interesting points of view on the
models.  In particular, it appears to enable one to define in
a transparent way non local observables providing a realization
of the algebra 
introduced by 't Hoooft in his work on quark confinement
\cite{tHooft,Fucito,Martellini1,Cattaneo}.

A natural question to be asked is whether the formulations
(\ref{BFYMAction}) and (\ref{BFYMetaAction}) of Yang-Mills theory
possess the same perturbative quantum properties as the standard
formulation.  This question has recently been answered in the
affirmative in the case of three spacetime dimensions \cite{Accardi},
where the theory is power-counting (super)renormalizable and anomaly-free for
any (simple) gauge group\footnote{For simplicity,
I assume throughout that the gauge group is simple.  This
excludes candidate anomalies of the form ``gauge invariant
polynomial" times ``abelian ghost", which could otherwise
arise in all formulations.  The results on the perturbative
equivalence of (\ref{YMAction}),
(\ref{BFYMAction}) and (\ref{BFYMetaAction}) are 
unaffected by this assumption.}.  This result was obtained in
\cite{Accardi} by entirely working out {\it ab initio}
the BRST cohomology at ghost numbers one (anomalies) and
zero (counterterms).  The purpose of this note is to point out that this
calculation can be completely shortcut by using the cohomological theorems
established in \cite{BBH1}.  Perturbative equivalence of the three
different formulations is direct and holds in any number
of dimensions.  In particular, (\ref{BFYMAction}) and (\ref{BFYMetaAction})
are power-counting renormalizable in four dimensions and there is
only one candidate anomaly, the Adler-Bardeen-Jackiw anomaly
\cite{PRL} (when the gauge group admits a non-zero $d_{abc}$-symbol).
In higher dimensions, the theory is not power-counting renormalizable
but renormalizable in the ``modern sense" 
\cite{Weinberg1,GomisWeinberg,Weinberg2} in either formulation.

To prove that the structure of the counterterms and of the
anomalies is the same in all three formulations, I shall follow the algebraic
approach to renormalization, based on the BRST symmetry 
\cite{BRS,Tyutin,Piguet}.  In that approach, the counterterms and
the anomalies are respectively described by the cohomological
groups $H^0(s\vert d)$ and $H^1(s \vert d)$.  These groups are
defined as the quotient spaces of the mod $d$ BRST cocycles, which
are the solutions of the Wess-Zumino consistency
condition
\bb
sa+db = 0
\label{WZ}
\ee
(at ghost number 0 and 1, respectively), modulo the mod $d$
BRST coboundaries $sm + dn$.  Here, $s$ is the BRST differential
acting in the standard manner on the fields, ghosts and antifields
(= sources for the BRST variations of the fields), while $d$ is
the spacetime exterior derivative.

The problem is thus to show that the cohomological groups $H^k(s \vert d)$
($k=0,1$) are isomorphic in all three formulations.  I first
show that this is the case for the formulations (\ref{BFYMetaAction})
and (\ref{BFYMAction}).

If one makes in (\ref{BFYMetaAction}) the change of
variables $B^a_{\mu_1 \dots \mu_{n-2}} \rightarrow
{\tilde B}^a_{\mu_1 \dots \mu_{n-2}}$, where 
${\tilde B}^a_{\mu_1 \dots \mu_{n-2}}$
is defined by (\ref{B}) and is invariant under the topological symmetry
(\ref{Gauge2}), one finds that the action (\ref{BFYMetaAction})
becomes exactly (\ref{BFYMAction}) with ${\tilde B}$ in 
place of $B$,
\bqn
S_{BFYM\eta}[A, B,\eta] &=&
\int d^n x \epsilon^{\mu_1 \dots \mu_n} B^a_{\mu_1 \dots \mu_{n-2}}
F_{a \mu_{n-1} \mu_n} \nonumber \\
&-&  2 (n-2)! g^2 \int d^n x
B^a_{\mu_1 \dots \mu_{n-2}}  B_a^{\mu_1 \dots \mu_{n-2}}.
\label{BFYMetaActionBis}
\eqn
Here, I have dropped the tilda on $B^a_{\mu_1 \dots \mu_{n-2}}$ and
I have discarded a surface term 
($\epsilon^{\mu_1 \dots \mu_n} D_{[\mu_1} \eta^a
_{\mu_2 \dots \mu_{n-2}]} F_{a \mu_{n-1} \mu_n}$ is a
total divergence because of the Bianchi identity).
Thus, in terms of these new variables, the field 
$\eta^a_{\mu_1 \dots \mu_{n-2}}$ does not appear in the action, and the
topological symmetry acts only on $\eta^a_{\mu_1 \dots \mu_{n-2}}$ 
as a shift symmetry,
$\eta^a_{\mu_1 \dots \mu_{n-2}} \rightarrow
\eta^a_{\mu_1 \dots \mu_{n-2}} + \Lambda^a_{\mu_1 \dots \mu_{n-2}} $.
Accordingly, a mere application of the theorems in section 15
of \cite{BBH1} shows that the cohomological groups $H^k(s \vert d)$
of  (\ref{BFYMetaAction})
and (\ref{BFYMAction}) are isomorphic.  I shall not
repeat the proof of the theorem here, but simply recall that it relies on
the fact that in any class of $H^k(s \vert d)$ for
(\ref{BFYMetaAction}), one can find a representative that does not
depend on $\eta^a_{\mu_1 \dots \mu_{n-2}}$, the
shift symmetry ghosts and the corresponding antifields.  This is because
these variables form contractible pairs that drop both from the
BRST cohomology and the mod $d$ BRST cohomology (\cite{BBH1}, page 85).

I now turn to the equivalence of (\ref{BFYMAction}) and
(\ref{YMAction}).  As stated above, one goes from
(\ref{BFYMAction}) to (\ref{YMAction}) by eliminating the auxiliary fields 
$B^a_{\mu_1 \dots \mu_{n-2}}$.  Now, the elimination of auxiliary
fields in the BRST-antifield formalism has been
investigated in \cite{Henneaux,BBH1}, where it has been shown that 
two formulations of the same theory differing only in their
auxiliary field content have isomorphic cohomology groups
$H^k(s \vert d)$ (\cite{BBH1}, Theorem 15.1).
The Yang-Mills example with field strength treated as
independent auxiliary variables was actually precisely considered
in \cite{BBH1}.  Since this result holds in particular for
$k=0$ and $k=1$, the quantum theories based on
(\ref{BFYMAction}) and (\ref{YMAction}) have equivalent
anomalies and counterterms.
Note that the auxiliary field dependence has the form of
Eq. (15.18) of \cite{BBH1}, which shows that polynomiality is
preserved under their elimination.  One may take the same
representatives of $H^k(s \vert d)$ before or after the
auxiliary fields are eliminated.

It follows that the structure of the anomalies and of the counterterms
is identical in the BF formulation (with or without
the topological symmetry exhibited) and in the standard second order 
formulation of Yang-Mills theory.

I close this letter by recalling what $H^0(s \vert d)$ and
$H^1(s \vert d)$ are in 3 and 4 dimensions.  First, as
shown in \cite{PRL,BBH2}, one may find representatives in
$H^0(s \vert d)$  and $H^1(s \vert d)$ that do not involve the
antifields (sources).  The cohomology is thus 
effectively reduced to the
antifield-independent cohomology described in
\cite{DuboisViolette1,Brandt,DuboisViolette}.
In three dimensions, there is no candidate anomaly,
$H^1(s \vert d) = 0$.  By contrast, 
$H^0(s \vert d)$ does not vanish and contains all the
gauge invariant operators, as well as the Chern-Simons term
$tr(AdA + (2/3) A^3)$, which is gauge invariant
only up to a total derivative and which is thus
associated with a non trivial descent.  These results
hold true without imposing any {\it a priori}
dimensionality condition on the cocycles and are
therefore also useful in the analysis of the
renormalization of (local or integrated) gauge invariant operators
of arbitrary dimensionality.
If one restricts oneself to the case of the action, as
in \cite{Accardi}, then power counting selects just two elements
in the above list, namely $F^a_{\mu \nu} F_a^{\mu \nu}$
and the Chern-Simons term.  This is in complete agreement with 
\cite{Accardi}.

In four dimensions, there is only one candidate
anomaly, the Adler-Bardeen-Jackiw anomaly,
\bb
 tr\left\{ C\left[ dAdA+
 \frac 12\left( AdAA-A^2dA-dAA^2\right) \right] \right\},
 \label{AdlerBardeen}
\ee
which does not vanish for non ``anomaly-safe" groups.  Here,
$C^a$ are the Yang-Mills ghosts.  But there
is no Chern-Simons term and $H^0(s \vert d)$ contains
only the strictly gauge invariant operators.  Power
counting then singles out $F^a_{\mu \nu} F_a^{\mu \nu}$
as the sole dimension 4 operator.

To conclude, I have illustrated in this letter the usefulness
of the cohomological theorems demonstrated in \cite{BBH1},
which imply straightforwardly that  the BFYM and YM theories
have equivalent candidate anomalies and  equivalent
counterterms.  In particular, equivalence holds in
three dimensions, as established in \cite{Accardi},
but also in four dimensions, a case not covered in
previous investigations.  Similarly, the renormalization of
gauge invariant (local
or integrated) operators presents identical features
in  the BFYM and YM formulations.

\end{document}